# Experimental investigation and scale analysis on melting of salty ice in a 3D-printed cavity filled with porous media

Li Xiaotian[1], WangYuming[2], Yang Wei[2], Yao wei[1, *]

1. Department of Space Exploration, Qianxuesen Laborotary of Space science and Technology, China Academy of Space science and Technology, Beijing 100094, China
2. Department o f Energy and Dynamics, College of Engineering, Peking University, Beijing 100871, China

## Abstract

While significant interests have been devoted to the double diffusive phase change heat transfer of binary solutions, the understanding on the melting heat transfer of salty ice in porous media is still imcomplete. This work aims to explore the melting heat transfer characteristics of salty ice in a square cavity filled with homogeneous porous media subjected to lateral heating. In order to facilitate the visualization of melting dynamics in a uniform porous media, a three-dimensional (3D) printed transparent cavity filled with porous matrix was manufactured, featuring an open upper end. Aqueous solutions of sodium chloride at concentration both higher and lower than the eutectic concentration were investigated. Lateral heating experiments with constant heat flux were conducted in a subzero temperature environment. The effect of heating power, initial concentrations and different sizes of porous matrix was investigated. Distinct phenomena were observed based on the concentration of the aqueous solution. Specifically, the concentration lower than the eutectic concentration results in an upward bending melting interface. Whereas the concentration surpassing the eutectic concentration resulted in a downward bending melting interface. In the presence of salt, a melting convex appears at the bottom of the melting interface and manifests during the whole melting process. Notably, smaller matrix enables faster melting rate. Furthermore, three distinct stages were revealed in our experiments, characterized as conduction, mixture of conduction and ultimately convection stage. The correlations for Nu and the average propagation rate of mushy/liquid interface were established by order of magnitude analysis and fit well with the experiments.

*. *Corresponding author*

*Yao Wei, 1. Department of Space exploration, Qianxuesen Laborotary of Space science and Technology, Chinese Academy of Space science and Technology, Beijing 100094, China, Mail.yaowei1972@hotmail.com*

# Introduction

Double diffusive convective heat transfter occurs in various nature processes and finds considerable applications in engineering contexts. The complication of heat transfer mechnisms during double diffusive convection are governed by a combination of temperature and concentration gradients within fluids. Researches and advancements on double diffusive convective heat transfer have been the interests of scholary communities for decacdes (Bennacer et al., 2001; Bourich et al., 2004; Mohamad and Bennacer, 2001; Trevisan and Bejan, 1990, 1985). Notably, a particular facet namely the phase-change double diffusive convective heat transfer has attracted significant attentions recently. The phase-change double diffusive heat transfer holds importance in diverse fields such as environmental science (e.g., solar ponds), oceanography (e.g., sea ice melting), geology (e.g., permafrost melting), astrophysics (e.g., extraterritial planets climate), semiconductor processing (e.g., crystal crystallisation), energy storage (e.g., phase change material). Extensive investigations have been devoted to both the fundamental aspects and application-oriented problems prevail to the double diffusion convective phase-change heat transfer.

Double-diffusion convection arising from melting or solidification of binary aqueous solutions without porous media has been extensively studied. Chen et al. (1971) studied the behavior of the flow field induced by lateral heating in a stably stratified fluid of constant gradient, and showed that the celluar convection was induced by the instability of fluid. Choi et al. (1980) experimentally studied the solidification of

$Na_2CO_3$ in various configurations, and found the occurrence of finger instability when cooling from bottom. Additionally, the opposing gradients of solutions resulted in the emergence of small double-diffusive layering. Thompson and Szekely (1988) explored the solidifacation of solutions from the lateral wall. A compositional and density stratification was discoverd during solidification. Christenson and Incropera (1989) experimentally investigated the lateral solidifications of solutions, revealing that the solidification rates, localized remelting, and macroscopic solute redistribution were influenced by the solutally driven flows. Marcoscopic mathmatical model related to the transport process was reviewed by Viskanta (1990) while Voller and Brent (1990) modelled the mushy region of a binary solidification system. Nishimura and Imoto (1993) reported the solidification of $NH_4Cl$-$H_2O$ in a cavity cooled laterally. Double diffusive cells with nearly constant concentration were identified within the liquid region during solidification and controlled by the buoyancy ratio and thermal Rayleigh number.

Cao et al. (1989) reported the solidification of aqueous sodium chloride solutions within a densely packed bed of glass spheres. Remelting phenomena were observed in the mush zone. They found that the enriched and stratified fluid layers appeared at the bottom of the fluid. The intensity of natural convection within this system was governed by the Darcy number and modified Rayleigh number. Matsumoto et al. (1993) studied the solidification of porous media saturated with aqueous solutions in a rectangular cell and characterized the solidification process both experimentally and numerically. The solidification rate is slow at the top and bottom due to the combined effect of thermal and salt buoyancy. The permeability coefficient in the mushy zone could be approximately expressed by the exponential formula for the liquid phase rate. Choi and Viskanta (1993) studied the solidification of binary alloys (water and ammonium chloride) in the saturated state of a spherical packed bed. Remelting was observed at both solid/mushy and mushy/liquid interfaces. Comparison of the experimental results with a simple analytical model that ignores the presence of convection reveals reasonable success in the solid and slurry regions. Song et al. (1993) experimentally and theoretically investigated the upward solidification of a porous medium saturated with a binary solution. An analytical model was proposed and predictions were compared with experimental data with good agreement. An "over-enrichment" phenomenon was also observed in the viscous region away from the liquid phase. Song et al. (2001) studied the experimental and theoretical investigation of lateral freezing of non-uniform porous media saturated with saline solution, which deepened the understanding of solute redistribution during solid/liquid phase transition. The presence of a porous structure significantly alters the distribution and growth kinetics of the primary solid phase. When the permeability of the porous matrix phase decreases, the predicted macroscopic devitrification decreases.

However, the melting heat transfer of binary mixture has received less attention. Beckermann and Viskanta (1990) experimentally investigated the melting of vertical ice into $NH_4Cl$-$H_2O$ solution and the melting of vertical $NH_4Cl$-$H_2O$ eutectic layer into water by laterally heating. It was found that in both cases the melting process results in the formation of a double diffusive layer in an initially homogeneous liquid. The experiments showed that relatively small changes in the thermal boundary conditions of melting could lead to large differences in the layering process and double diffusion in the liquid also had a significant effect on the melting process. Yin et al. (2000) investigated the development of layered structures, convective mixing of indium and gallium, bidirectional flow patterns, and solid-liquid interfaces. The presence of layering and multiple double convective layers in the binary Ga-In5 melt was observed in real time by X-ray visualization under transient horizontal temperature gradients. Multiple double diffusive layers were interspersed at the top of the melt layers. A close relationship between the double diffusive layers and the solid/liquid interface was identified. The interface is divided into several parts by characteristic morphological protrusion features.

The melting process of binary eutectics in porous media is also very important. For example, the melting of salty ice occurs widely in thawing of frozen soil, Mars exploration (Mellerowicz et al., 2022), thermal storage , etc. However, only one study on the melting of binary eutectics in porous media was identified through literature investigation. Ouestlai et al. (2008) used numerical methods to study the effect of density change on double diffusive convection of meltwater in a side-heated porous medium considering a nonlinear density expression (Oueslati et al., 2008). The author chose an initial temperature of 0°C and used initial concentrations of 5 and 10 wt%, at which point the solution did not freeze due to the lowering of the freezing point, so the conclusions of the article are questionable.

In order to better understand the mechanism of double diffusive convective heat transfer in the melting in low temperature environments, we conducted experiments with 3D printing technology to faciliate visualization. The natural convection and heat transfer characteristics of the melting process in transparent cavity with saturated porous matrix were investigated. The correlations of Nu number and average propogation speed of melting interface were established by the order of magnitude method for the hypoeutetic region.

## Experimental equipments and procedures

As shown in Fig.1, a transparent square container was fabricated with 3D printing method to facilitate direct visualization of the melting process. A desktop-scale experimental setup in the laboratory was built to simulate the subzero temperature environment of Mars. The transient evolution of melting interface was filmed by a Canon$^{TM}$ Legria HFR 86 video camera and the temperature data was measured by thermocouple and recorded by a digital data acquisition unit during the melting process. The cavity is 40 mm wide, 40 mm high, 2 mm thick, and has a 2 mm base for fixation. A rectangular array of small black cylinders was printed inside the square cavity with the integrated 3D printing technology to simulate the solid matrix in a porous medium. A cylindrical cavity with a diameter of 6 mm, a height of 40 mm, and a wall thickness of 1 mm is attached to the left side of the square cavity for mounting a cylindrical heater.

During the experiment, a metal heating rod with a diameter of 6 mm and a height of 40 mm was placed into the cavity, and the inside of the cavity was filled with thermally conductive silicone grease to reduce the contact thermal resistance between the heating rod and the wall. A schematic diagram of the test cell is shown in Figure 2.

Twenty-five type K thermocouples (wire diameter 0.21 mm) were taped to the outside of the square cavity to measure the transient temperature distribution of the square cavity shell.

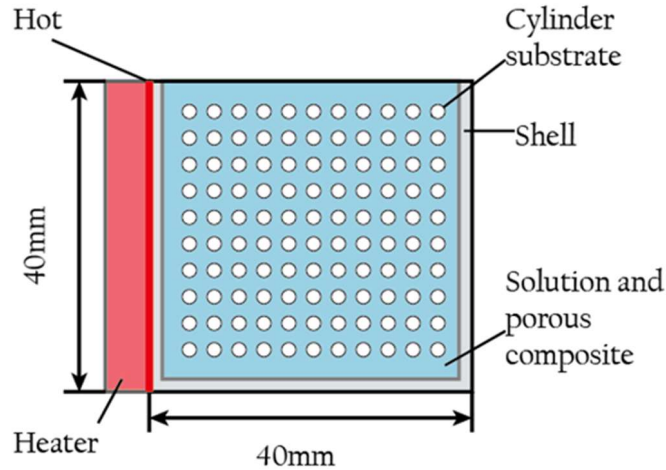

Figure 1 Schematic of the assembled cavity with soluiton saturated porous matrix

The entire cavity was fixed to an acrylic base by screws to eliminate the effect of tilting of the square chamber on convection. Deionized water was mixed with the desired NaCl in different ratios to obtain the desired salt concentration. The solution was heated to 80 °C to remove the bubbles from the water, and the degassed aqueous NaCl solution was carefully injected into the interior of the square cavity using a hypodermic syringe (5 mm diameter of the needle) to minimize the gas bubbles in the solution. Prior to the test, the solution-filled square chamber was frozen in a low-temperature refrigerator. Since the addition of NaCl lowers the freezing point of the aqueous solution, the refrigerator temperature was set to -40ºC during freezing to allow the solution to freeze completely and rapidly, which lasted for 10 minutes, to keep the solute as evenly as possible.

A Qingsheng™ QS2000 thermal cycling chamber was used to simulate the low temperature environment. The lowed temperature of the chamber is -40 ºC. A square chamber filled with the frozen solution was placed inside the chamber and the left wall of the chamber was heated by an electric cylinder heater controlled at a constant power level. Before starting the test, the hot/cold circulator was turned on for refrigeration, and after confirming that the temperature inside had stabilized to the desired low temperature by observing the display, the door of the circulator was opened. Then the square chamber, which had been fixed on the base, was taken out of the refrigerator and quickly put into the hot/cold circulator. Then the door of the circulator was closed. For the visualization test, the outer walls of the square chambers used were free of thermocouples to avoid interference with the filming. For temperature measurement tests, K-type thermocouples were uniformly attached to the outer wall of the square chamber. The thermocouple tail leads were routed through a small hole on the left side of the recirculation box and the hole was plugged with a rubber stopper. All type K thermocouples were calibrated with ice water at 0 °C, and a temperature data acquisition system was used to record temperature changes at predetermined time intervals (30 s). During the course of the experiments, when a melting liquid appeared, a trace amount of dry methyl blue powder was pushed into the cavity via a hypodermic syringe suspended above the left wall surface in the square cavity to stain the melting zone in order to better distinguish the solid-liquid interface. To perform the visualization experiments, a white LED planar light source was used to illuminate from behind the square cavity. During the visualization experiments, the front and back of the square chamber were covered with a heat shield, while the top and bottom of the chamber were in direct contact with cold air. Photographs were taken at 5 min intervals, and the heat shield was removed when taking the photographs, and then the front and back of the cavity were covered after taking the photographs in order to minimize the heat loss during the photographs. For temperature measurement experiments, the heat shield was kept over the square cavity. Figure 2 shows the optical picture taken without the heat shield.

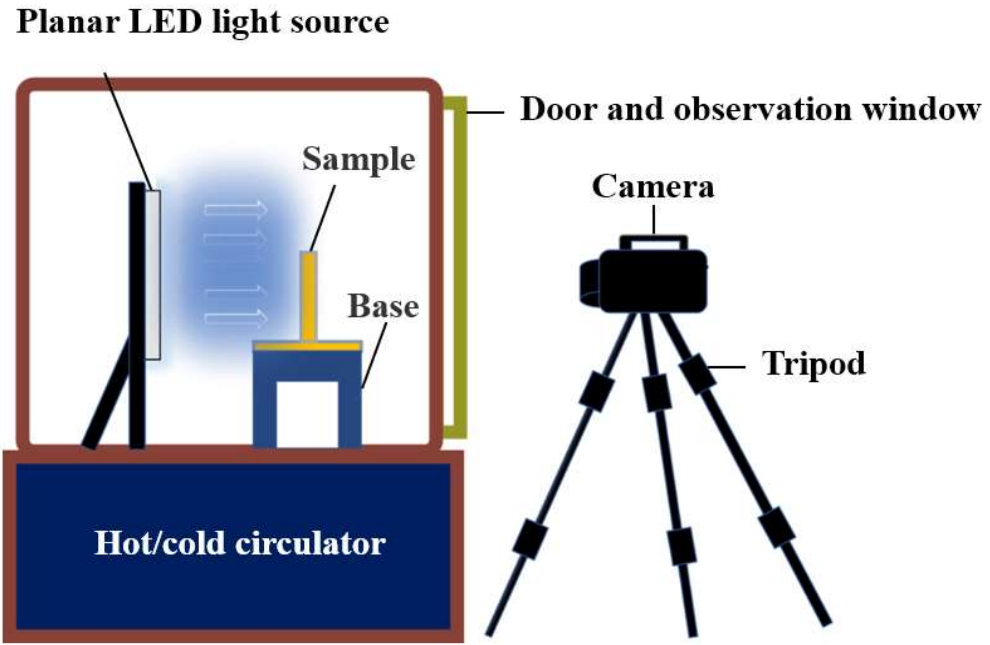

Figure 2 Schematic diagram of the test setup

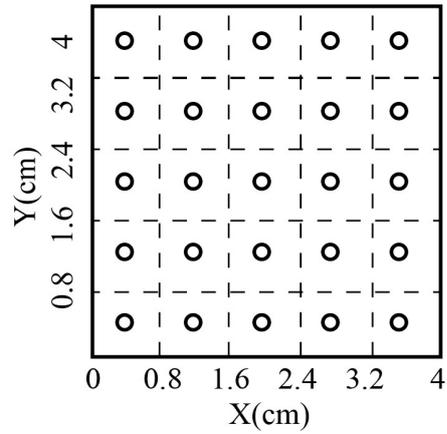

Figure 3 Schematic of thermocouple arrangement with a spatial separation of 0.8 centimeters between each consecutive thermocouple.

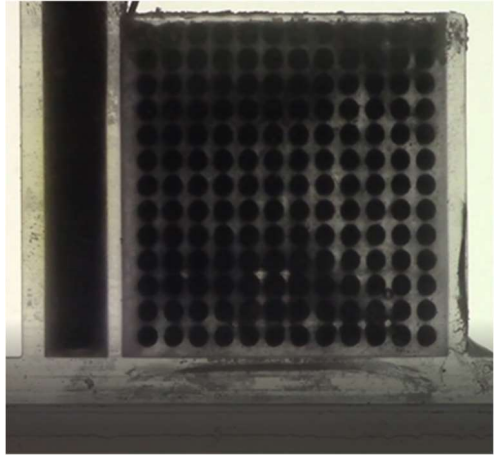

Figure 4 Phtograph of forzen sample

Six sets of experiments were carried out and the conditions are shown in Table 1. Experiments 1-3 are to study the effect of different heating powers on the melting process. Experiments 4-5 are to investigate the effect on the melting process by measuring the change in initial concentration, and Experiment 6 is to explore the effect of different porous matrix diameters on the melting process. Each set of experiments was performed with three repetitions of visualization tests and three repetitions of temperature measurements, respectively. In this case, Experiment 2 was used as a control group with 5 wt.% initial salt concentration, 5 W heating power and 2 mm cylinders. The eutectic point of the aqueous NaCl solution used was -22.4 °C and the eutectic concentration was 23.1 wt.%. In all experimental conditions, the ambient temperature was -30ºC.

Table 1 Summary of Ice Melt Experimental Conditions

| Working condition | Initial concentration | Heating power | Cylinder diameter |
|---|---|---|---|
| 1 | 5 wt.% | 3 W | 2 mm |
| 2 | 5 wt.% | 5 W | 2 mm |
| 3 | 5 wt.% | 7 W | 2 mm |
| 4 | 10 wt.% | 5 W | 2 mm |
| 5 | 25 wt.% | 5 W | 2 mm |
| 6 | 5 wt.% | 5 W | 1 mm |

# Experimental results and discussion

The photographs of the melting process of the binary mixture in the cavity is shown in Figure 5. The lighter part is the liquid region and the darker part is the mushy region, while the darkest part is the frozen region. The mushy/solid interface was identified as melting interface in this article. In the initial stage, the melting interface propogate faster at the center. This is due to that the top and bottom of the square cavity are in conctact with the cold environment exchanging heat with it, thus the melting rate is lower. At 5 minutes, the melting interface is approximately parallel to the left wall, mainly because the heat transfer mechanism is conduction. At 10 minutes, the melting region at the top of the square cavity grows faster, indicating that thermal convection has developed in the upper part of the melting region. The fluid ajacent to the left wall rises along the heated wall and accumulates at the top of the square cavity, thus accelerating the melting in the upper part. At the bottom of the square cavity, the melting also starts to accelerate, forming a nose protruding to the right, which is mainly due to the redistribution of the solutes in the liquid zone adjoining the melting interaface. When the eutectic mixture melts into solution, the solute depletion occurs in this region due to the difference in solubility of NaCl in the liquid and solid phase. Solutes precipitate out of the ice into the mushy zone(Brimblecombe et al., 1987; Kumar et al., 2007; Terwilliger and Dizon, 1970), thus increasing the concentration and resulted iin the denser liquid near the melting interface. The denser liquid sinks to the bottom and lowers the freezing point, thus accelerating the melting of the frozen solution at the bottom. A stagant zone was build at bottom and double diffusion layer appeared between warmer, less salty liquid at the upper section and colder, more salty liquid at the bottom. The mechnism is different from Yin et al. (2001). In their melting experiments, the unmelted In elements snow down the bottom from the interface and increase the thermal conductivity at the bottom, which leads to an accelerated melting rate. Unlike Beckerman et al. (1990) as well, their experiments involved eutectics dissolving into fresh water, increasing the concentration of neighboring fresh water, which increased the density of the water and snow down to the bottom, thus increasing the melt rate at the bottom.

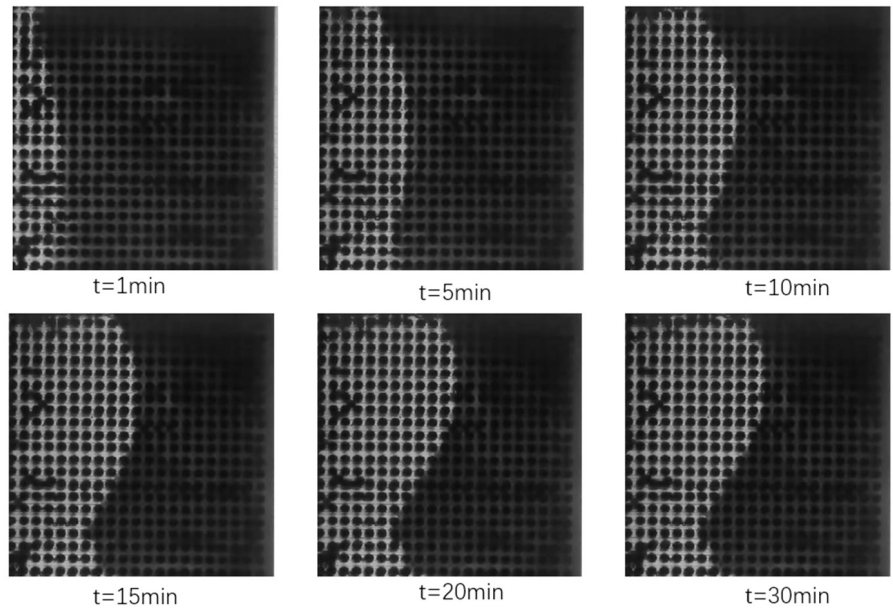

Fig. 5 Evolution of the melting interface for the heating power of 5 W, the initial salt content of 5 wt.%, the porous matrix with small cylinders of 1 mm diameter and 0.5 mm spacing, and an ambient temperature of -30 °C. (In the figure, the light-colored part is the melting region, the dark-colored part is the frozen region, and the melting interface is the intersection line between the light-colored and dark-colored parts, and the subsequent photographs are the same)

As shown in Fig. 6, the melting interface evolution is plotted for the heating power of 5 W, the initial salt content of 5 wt.%, and the small cylindrical diameter of 2 mm, at the ambient temperature of -30 °C. The melting is slower in the initial moments in the case of the 2 mm matrix than in the case of a 1mm matrix, and is still predominantly by thermal conduction at 5 minutes. Although the spacing is the same, the larger diameter results in a smaller number of matrices per unit volume, and therefore allows for a larger pore volume and greater porosity for the development of convection. However, in terms of heat transfer, the thermal conductivity of the matrix decreases due to the increase in matrix porosity, therefore its heat transfer capacity will decrease and the melting rate will be slower. The increase in porosity increases the convective strength, but the increase in thermal convection is relatively more pronounced. The effect of the enhancement of thermal convection and decrease in thermal conductivity on the melting rate, on the other hand, needs to be analyzed quantitatively and will be discussed further in the subsequent parts of the article.

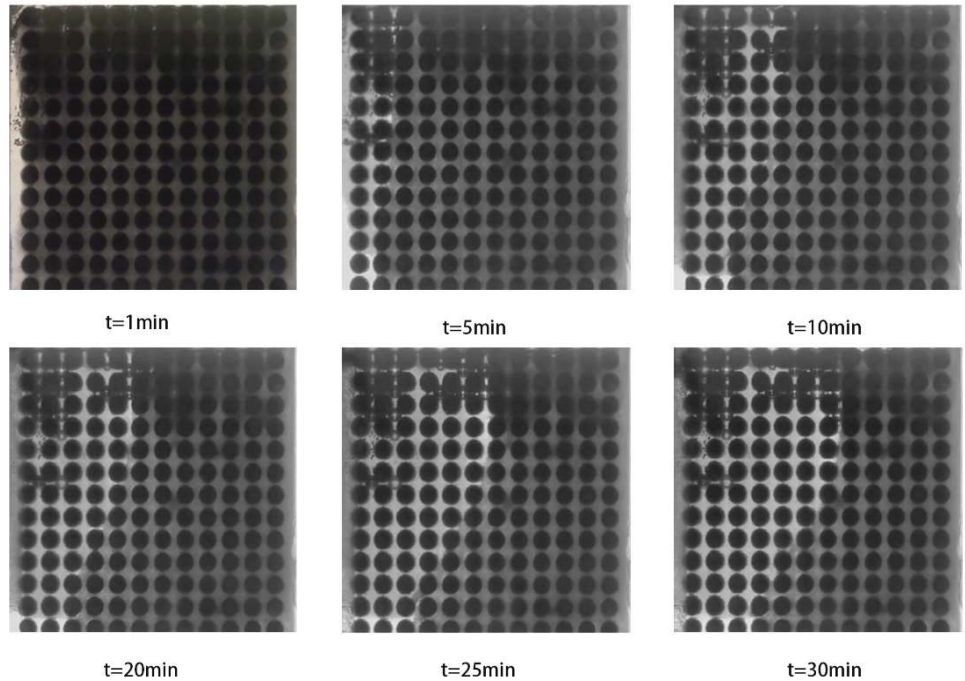

Fig. 6 Melting interface evolution for the heating power of 5 W, the initial salt content of 5 wt.%, and the porous matrix with small cylinders with diameter of 2 mm at the ambient temperature of -30 °C

As shown in Fig. 7, the evolution of the melting interface is photographed for a heating power of 5 W, an initial salt content of 10 wt.%, and a porous matrix cylinder with a diameter of 2 mm and a spacing of 0.5 mm. Compared with the 5 wt.% concentration, the melting rate increases significantly, with a more pronounced interfacial bending observed at nearly 10 min. This phenomenon is attributed to the inverse relationship between concentration and freezing points. The higher concentrations lowered freezing points and accelerated the melting rate. With the heat conduction mechanism, the melting interface propogated to the right for a longer distance due to the faster melting rate. During this stage, heat convection mechanism is not yet fully developed. The upper section of the melting interface bends to right rather than to left. The melting rate is higher at a concentration of 10 wt.% as opposed to the 5 wt.%. This augmented melting area intensified the thermal convection against the heat loss between the upper section of the square cavity and its surroundings. Thus, the melting rate at the upper scetion was promoted. For the 5 wt.% case, on the other hand, the cumulative effect of heat at the top due to thermal convection is weaker than the heat loss by convective heat exchange between the top and the surrounding, so the temperature at the top decreases and the temperature in the middle of the interface is higher. Correspondingly, the rate of motion of the top interface is slower than that of the middle, which is manifested by the melting interface bending to the left at the top.

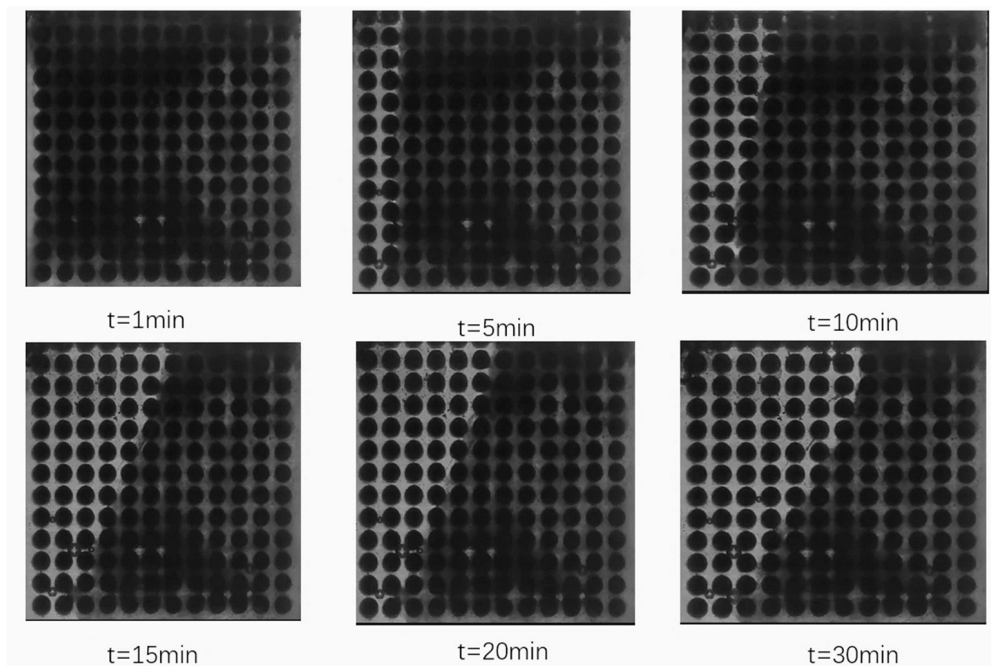

Fig. 7 Evolution of the melting interface for the heating power of 5 W, the initial salt content of 10 wt.%, small cylinders of porous matrix with a diameter of 2 mm and a spacing of 0.5 mm, and the ambient temperature of -30 °C.

As shown in Figure 8, the photographs of the melting process for Test 5 are presented. In contrast to the 5 wt.% and 10 wt.% cases, the melting interface bends downward to the right. It is simply because at this concentration the solubility already exceeds the eutectic concentration, salt precipitated and settled to the bottom of the square cavity once the metling occured. When the melting started, the salt concentration at the bottom was higher, which greatly reduced the melting point, so the melting rate in the lower part was much higher than in the upper part. This is similar to Beckerman et al. (1990). With the increase of the melting region, the thermal convection is enhanced. While the effect of salt concentration gradient on the flow pattern is much larger than that of thermal convection. Due to the influence of the salt concentration gradient in the vertical direction, the melting interface continues to bend to the lower right, while the area influenced by thermal convection is small. So, the temperature at the top is very low, resulting in lower melting rate. With the enhancement of thermal convection, the effect caused by the salt concentration gradient is partially canceled at the upper section, where the degree of bending gradually decreases and an inflectioin appears as time elpased.

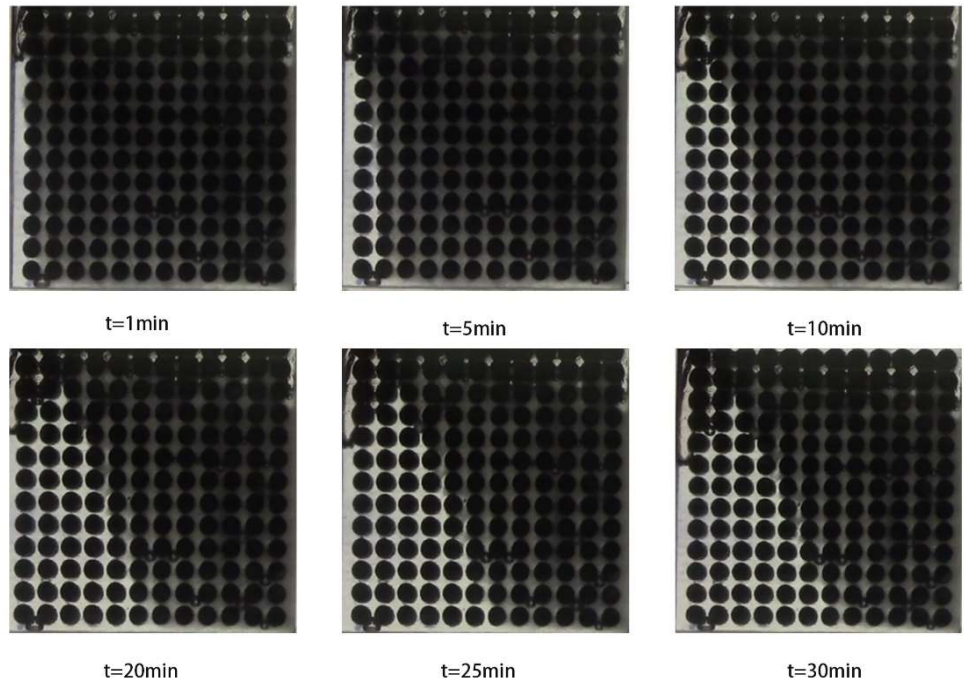

Fig. 8 Evolution of the melting interface for the power of 5 W, the initial salt concentration of 25 wt.%, small cylinders of porous matrix with a diameter of 2 mm and the spacing of 0.5 mm, and an ambient temperature of -30 °C.

In order to further quantitatively analyze the melting rate for different working conditions, the liquid fraction as a function of time in the cavity under different working conditions was analysed using matlab and ImageJ software. The melting rates were compared for different initial salt concentrations, porous media matrix sizes, and heating powers, respectively.

As shown in Fig. 9, the variation of liquid fraction with time for different initial salt concentrations is shown. Overall, the melting becomes faster as the salt concentration increases. The reason for this phenomenon is that the melting point decreases as the concentration increases, boosting the melting rate. In addition, the melting rate gradually slows down with time. The first turning point of the curves occurs after 5 min, implying the development of convection. A second turning point occurs after 15min, indicating heat equilibrium with ambient is gradually built at this point. Between 15min and 30min, the liquid fraction changes little, indicating that a steady state is almost reached at this point. Comparatively speaking, the melting curves of 5wt.% and 10wt.% mixture have similar trends, which can be almost repeatible by merely curve translation, because their flow patterns are dominated by heat convection and the melting interface is deflected to the upper right. The flow pattern of melting of 25 wt.% mixture is different from the previous two. The melting interface is deflected downward to the right, and thermal convection is limited due to the solutal gradient, so the shape of the liquid fraction curves is also different.

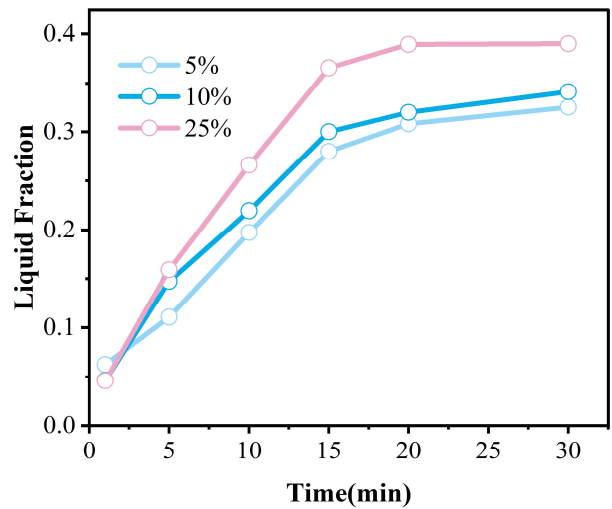

Fig. 9 Variation of liquid phase rate with time in the square cavity for different initial salt concentrations

As shown in Fig. 10, the liquid fraction curves are shown for different diameters of porous matrices. There lies an inverse relationship between the diameter of the porous matrix and liquid fraction. As the diameter decreases, the volume of each pore decreases, resulting in a higher quantity of porous matrices per unit volume and leading to a reduction in porosity. However, the thermal conductivity is increased. The decrease in porosity constrains the developement of thermal convection, while the consequential increase in thermal conductivity exerts a more crucial influence on the melting rate. Consequently, the net effect is an enhancement in the overall liquid fraction, due to the reduction of the diameter of the porous matrix. This complicate interplay between porosity, thermal conductivity, and melting rate unfolds the dynamics governing heat transfer mechnism within the porous matrix.

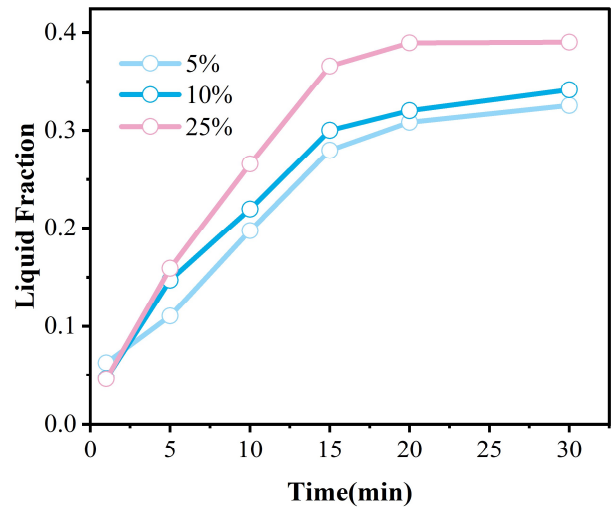

Figure 10 Melting rate for different porous matrix diameters

The liquid fraction profiles for various heating powers are shown in Fig. 11. As the heating power increases, the melting rate grows. The higher the power, the faster to reach thermal equilibrium. This is because a faster melting rate leads to an earlier initiation of the thermal convection. After the fully developement of thermal convection, the thermal equilibrium is built (Jany and Bejan, 1988). In our experiments, due to the low ambient temperature and heating power level, after reaching thermal equilibrium, the melting rate decreases gradually and the melting interface couldn't make contact with the right-side wall.

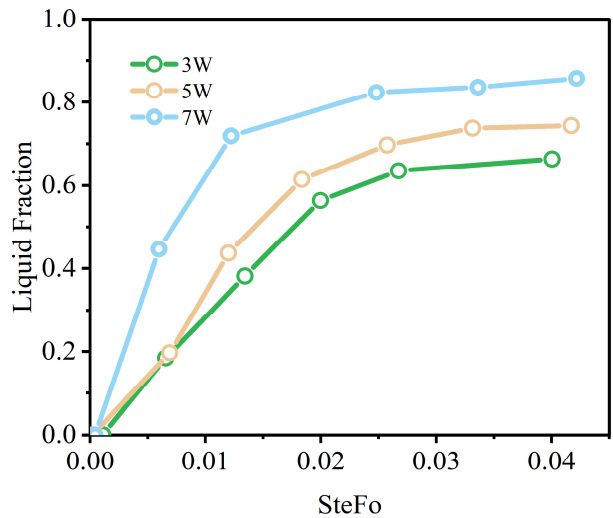

Fig. 11 Variation of liquid phase rate with time for different heating powers

## Order of magnitude analysis

The visualized experimental results indicate that the melting process could be devided into three stages, as shown in Fig. 12, which differs from four stages of Jany et al. (1988) due to the melting interface couldn't make the contact with the right wall in our experiments. Heat conduction dominates the melting process in the early stage of melting. During conduction stage, heat flow is perpendicular to the hot wall and the melting front is almost parallel to the heated wall. Then the thickness of the liquid region increased and natural convection was developed. As a result, the heat flux along the heated wall becomes inhomogeneous, which deforms the melt interface, indicating the mixture heat transfer stage starts. In the final stage, natural convection develops over time and dominates the phase change process. In total, natural convection is affected by both the thermal and solutal gradients, which influences the heat transfer mechanism and thus the melting process. In order to further clarify the heat transfer mechanisms, a theoretical analysis based on order of magnitude analysis is performed on the hypoeutectic side.

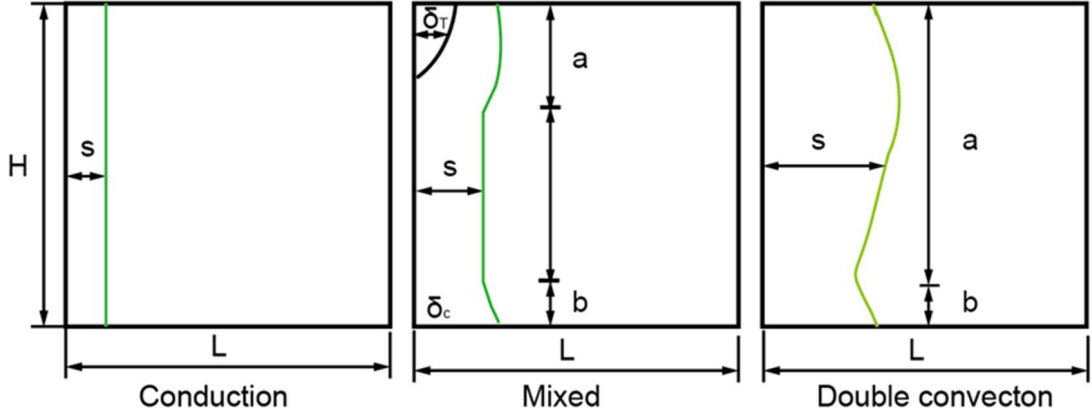

Figure 12 Schematic diagram of solid-liquid interfacial evolution, with the liquid phase on the left of the green line

**(1) Heat conduction stage**

As shown in Fig. 12 (a), the melt front is almost parallel the heated wall. The order of s is comparable with H. By applying the law of conservation of energy to the liquid-phase region, the density of heat flow introduced through the left wall is of the same order of magnitude as the enthalpy change of melting at the interface. According to the analysis, it is obtained

$$\underbrace{\frac{k(T_w-T_m)}{s}}_{\text{heat conduction}} H \sim \underbrace{\frac{\rho h_m s}{t}}_{\text{heat of fusion}} H$$

where $h_m$ is the enthalpy of melting, $k$ is the thermal conductivity, $h_0$ is the forced convective heat transfer coefficient with the environment, and $T_{amb}$ is the temperature of the ambient. The order of magnitude of the average distance of the melting interface propogated from the heated wall can be derived as:

$$s \sim H(\theta)^{1/2}$$

where $\theta = Ste\ Fo$,

$$Nu_0 = \frac{h_0 L}{k}$$

$$Ste = \frac{C(T_w-T_m)}{h_m}$$

$$Fo = \frac{\alpha t}{H^2}$$

From the eqation, it is known that the distance of the melting interface from the left heating wall shows a relationship with the height of the square cavity, the heating time and the specific heat capacity of the fluid itself, the thermal diffusivity and the enthalpy of melting. The solute affects the thermophysical properties of water to some extent, and therefore indirectly affects the rate of motion of the melting interface at this stage. Under constant heat flow density conditions, $T_w - T_m$ is unknown. $(q''s)/c \sim (T_w - T_m)$, where $q''$ is the heat flow density.

The average Nu number of the heated wall is:

$$Nu = -\frac{1}{T_w - T_m} \int_0^L \left.\frac{\partial T}{\partial x}\right|_{x=0} dy$$

Substituting $dT \sim T_w - T_m$, $dx \sim s$, gives

$$Nu \sim \frac{H}{s} \sim H\theta^{-\frac{1}{2}}$$

**(2) Mixed heat transfer stage**

As the melting continues, thermal convection was activated and intensitified in the upper section of the cavity as the liquid region increases, accelerating the melting process. As a result, the liquid region in the upper section becomes wider with a height of a, whereas in the lower section with a height of H-a, is still dominated by thermal conduction, including the stagnant region at the bottom. At the mixed heat transfer stage, the heat flux is introduced along the heated wall and is in equilibrium with the fluid heat transfer. The fluid heat transfer can be divided into two sections: the upper section dominated by thermal convection heat transfer and the lower section dominated by heat conduction.

Within the convective region, which can be divided into the main flow region and the boundary layer, the changes of temperature and concentration mainly occur in the boundary layer, so $s \sim \delta_T, H - a \sim H$.

In the initial stage, heat conduction is predominant, so according to the conservation of energy:

$$\frac{\partial T}{\partial t} \sim \alpha \frac{\partial T}{\partial x^2}$$

Applying substitution yields

$$\frac{\Delta T}{t} \sim \alpha \frac{\Delta T}{\delta_T^2}$$

The thickness of thermal boundary layer can be found as

$$\delta_T \sim (\alpha t)^{\frac{1}{2}}$$

According to mass conservation

$$\frac{\partial u}{\partial x} + \frac{\partial v}{\partial y} = 0$$

where u is the velocity in x-direction, v is the velocity in y-direction. Applying the order of magnitude analysis yields

$$\frac{u}{\delta_T} \sim \frac{v}{H}$$

The x-direction omentum equation is

$$0 = -\frac{\partial p}{\partial x} - \frac{u}{Da}$$

The y-direction momentum equation can be rewritten as

$$0 = -\frac{\partial P}{\partial y} - \frac{v}{Da} + \frac{g\beta_c \Delta T \theta L}{\nu v} + \frac{g\beta_T \Delta C \phi L}{\nu v}$$

Taking the partial differential ∂/∂y for the x-direction momentum equation, and the partial differential ∂/∂x for the y-direction momentum equation, and subtracting them yields

$$\frac{\partial v}{\partial x} - \frac{\partial u}{\partial y} = Da(\frac{g\beta_c \Delta T \theta L}{\nu v}\frac{\partial \theta}{\partial x} + \frac{g\beta_T \Delta C \phi L}{\nu v}\frac{\partial \phi}{\partial x})$$

According to the phase diagram of NaCl solution, in the hypoeutectic side, the liquidus line is approximated as a slash and the dimensionless melting point can be expressed by $\theta \sim -m_i \phi$, thus, leading to $\theta \sim \phi$, where $m_i$ is the slope of the liquidus line. The above equation can be rewritten as

$$v \sim Da(\frac{g\beta_c \Delta T \theta L}{\nu} - \frac{g\beta_T \Delta C \phi L}{\nu})c$$

When steady state is achieved, the buoyancy force is balanced by the Darcy frictional resistance. At this equilibrium point, the velocity of the fluid becomes constant and can be expressed as

$$v_{ref} \sim \frac{g\beta K(1-N)\Delta T}{\nu \delta_T}$$

The buoyancy force is determined by the density difference casused by the gradient of temperature and concentration. Upon entering the convection dominated stage, the heat conduction term in the energy equation can be neglected, thus

$$\sigma \frac{\Delta T}{t} \sim v \frac{\Delta T}{H}$$

From this we can find

$$t_f \sim \frac{\sigma H}{v}$$

At this point, the thermal boundary layer maintains nearly constant, so

$$\delta_T = H[(1-N)Ra_m]^{-\frac{1}{2}}$$

where N is the buoyancy ratio and $Ra_m$ is the Darcy Rayleigh number, the height of the convection layer is $a$, which leads to

$$\frac{\delta_T}{Z} \sim \left((1-N)Ra_{m,z}\right)^{-\frac{1}{2}}$$

where $Ra_{m,z} = (Ra_m z)/H$. At the bottom, the width $s$ and height are of the same order, so

$$z(1-N)Ra_{m,z} \sim H(\text{Ste } Fo)^{\frac{1}{2}}$$

Therefore, the order of z is

$$z \sim (1-N)H Ra_m \text{ Ste } Fo$$

The mixing heat transfer stage ends at when z equals H. Therefore, the mixing heat transfer phase ends at

$$\theta_1 \sim [(1-N)\ H Ra_m]^{-1}$$

According to the above analysis, the heat transfer mechanism consists of two parts: convection and conduction. The average Nu number of the heated wall is

$$Nu = \frac{H-z}{s} + \frac{z}{\delta_T}$$

Substituting into the above equation yields:

$$Nu \sim (SteFo)^{-\frac{1}{2}} + (1-N)Ra_m \text{Ste } Fo^{\frac{1}{2}}$$

**(3) Convection stage**

When the bottom of the thermal convection region touches the top of the salt convection region, we consider that convection has entered the fully developed stage after which $z = H$, thus

$$Nu \sim \frac{H}{\delta_T}$$

At the stage of full convective development, both the thermal boundary layer and the convective height remain constant, so the $Nu$ number remains constant

$$Nu \sim (1-N)Ra_m$$

Compared with Jany et al.'s study, we introduce the effect of salt and porous media on the melting process, which is reflected in the correlation equation by increasing the buoyancy ratio $N$ and the Darcy number $Da$, where $Da$ can be coupled into the Darcy Rayleigh number $Ra_m$.

## Comparison with tests

In order to verify the results of the theoretical analysis, we analyzed the average $Nu$ number of the heated wall and the average propogation speed of the melting interface under different working conditions. The formulas for data processing are as follows. The average heat transfer coefficient is

$$\bar{h} = \frac{Q_{heat}}{A(T_w - T_m)}$$

where $T_w$ is the average wall temperature, $T_m$ is the melting temperature, A is the heat transfer area, $Q_{heat}$ is the net heating power. Then the $\bar{Nu}$ caculated from the data is

$$\bar{Nu} = \frac{\bar{h}L}{k}$$

where $L$ is the characteristic length and $k$ is the thermal conductivity

$$q_{inter} = \frac{\rho_s h_{sl} dV}{dt}$$

where $q_{inter}$ is the heat transfer rate at mushy/re, $\rho_s$ is the solution density, $h_{sl}$ is the latent heat of melting, and $V$ is the liquid phase volume. Stefan number can be expressed as

$$Ste = \frac{C_p \left(\frac{qL}{k}\right)}{h_{sl}}$$

where $C_p$ is the specific heat capacity, $q$ is the input heat flux, $L$ is the characteristic length, and $h_{sl}$ is the latent heat of phase transition.

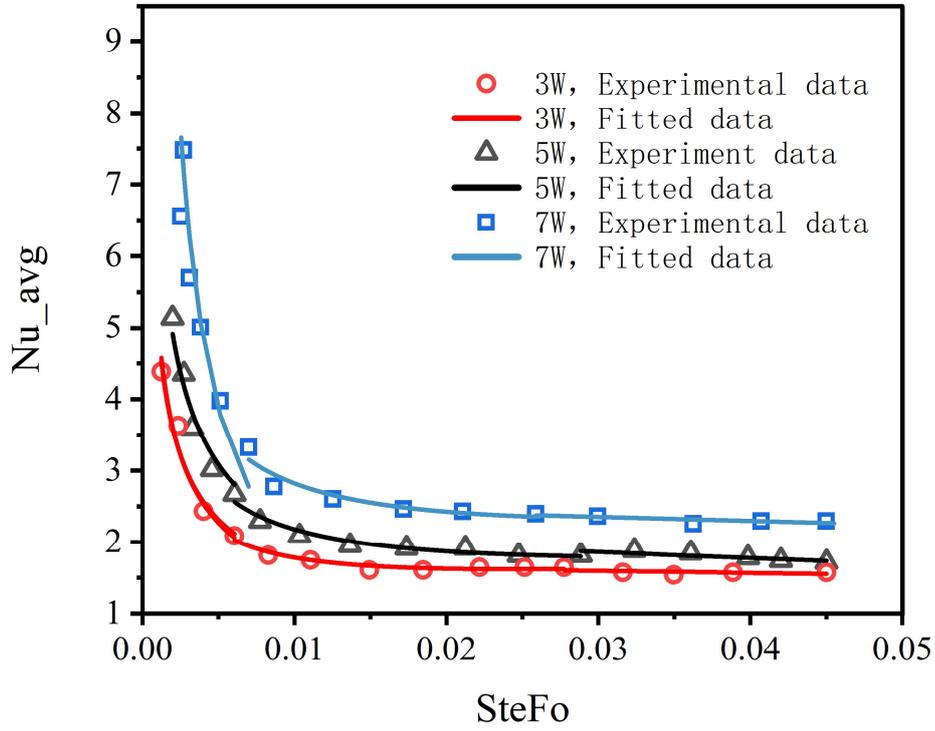

Fig. 13 Average Nu number of heated wall with dimensionless time

Table 2 Fitting Functional of $Nu$

| Power (W) | Conduction stage | Mixed heat transfer mechanism stage | Convection fully developed stage |
|---|---|---|---|
| 7 | $y = 0.126 * SteFo^{-0.5}$ ($0 < SteFo < 0.007$) | $y = 0.126 * SteFo^{-0.5} - 2.83 * SteF^{0.5}$ ($0.007 < SteFo < 0.026$) | $y = 0.10 * Ra_m^{-0.25}$ ($0.026 < SteFo < 0.04$) |
| 5 | $y = 0.086 * (SteFo)^{-0.5}$ ($0 < SteFo < 0.006$) | $y = 0.085 * (SteFo)^{-0.5} + 3.88 * SteF^{0.5}$ ($0.006 < SteFo < 0.029$) | $y = 0.080 * Ra_m^{-0.25}$ ($0.029 < SteFo < 0.04$) |
| 3 | $y = 0.063 * (SteFo)^{-0.5}$ ($0 < SteFo < 0.006$) | $y = 0.012 * (SteFo)^{-0.5} - 5.55 * SteF^{0.5}$ ($0.006 < SteFo < 0.028$) | $y = 0.078 * Ra_m^{-0.25}$ ($0.028 < SteFo < 0.04$) |

In order to verify the results of the order-of-magnitude analysis, we carried out experiments with input powers of 3W, 5W, and 7W, measured the temperature data and processed the relevant data according to the above equations. As is shown in Fig.13, the $Nu$ number increases as the input power increase. The three different power conditions share the same tendency, indicating that the flow pattern in the liquid phase region is similar. The change of $Nu$ can be roughly divided into three successive stages. At the first stage $Nu$ decreases rapidly, indicating the predominance of the heat conduction. The propogation of the melting interface to the right increases the heat transfer thermal resistance.

The second stage occurs when notable inflection point appears, indicating a transition in the heat transfer mechanism. Convective heat transfer is activated and evolves into a mixed convection mechanism alongside conduction. In the third stage, the $Nu$ number curve approaches to be parallel with the horizontal coordinate axis, exhibiting a nearly constant value independent of time. This implies the dominance of convective heat transfer approaching to a fully developed state, or a quasi-steady state. Stages of transformation were initially determined by comparing interfacial evolution diagrams during the melting process, followed by fitting each stage using the order-of-magnitude correlations. Notably, the average Nu number under each input power fits well with the experimental results, affirming the validity of the order-of-magnitude analysis. The expressions of the fitted function are detailed in Table 2.

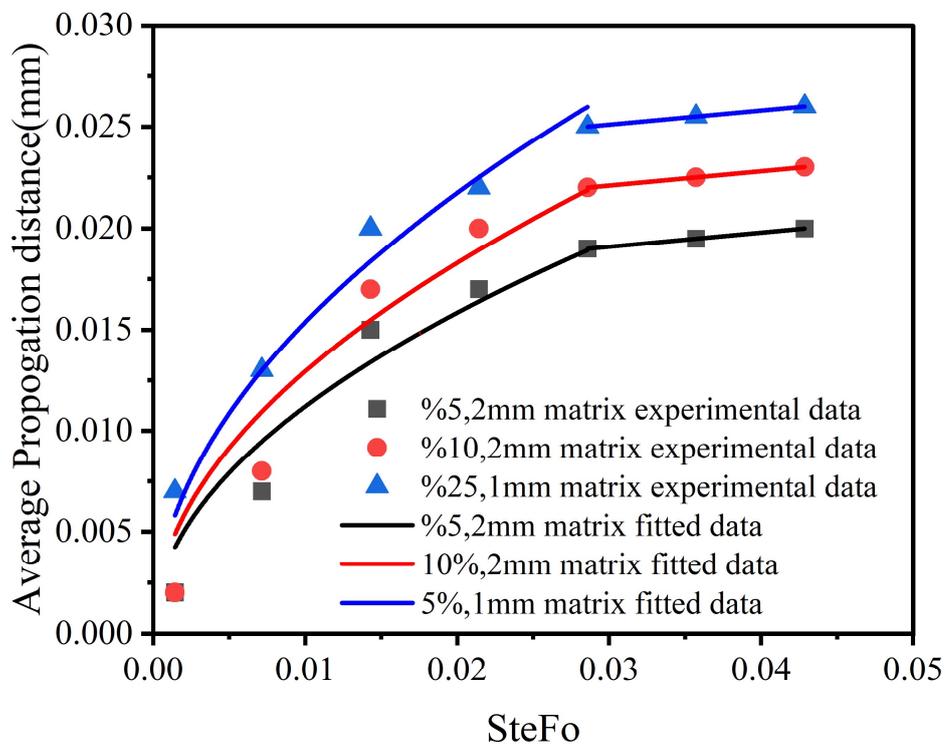

Figure 14 Plot of average melt interface travel distance against dimensionless time.

As shown in Fig. 14, the propogation distance of the melting interface is plotted as a function of SteFo for different concentrations and matrix diameters. Under identical porous media parameters, the increase in concentration leads to the increase of movement speed of the melting interface. While under identical concentration, the reduction in the diameter of the porous medium matrix results in the increased average movement rate of the melting interface. The variation of the average displaced distance of the melting interface can be divided into two stages. The first phase lasts for a about $L(SteFo)^{-\frac{1}{2}}$ until the end of the heat conduction. The correlations of the second phase is approximately $LSteFoRa^{-\frac{1}{4}}$. Both of the expressions show alliance with the experimental data.

## Conclusion

An experimental visualization of the melting process of a frozen binary solution within a square cavity subjected to lateral heating is performed. The experiments employ a 3D-printed transparent square cavity filled containing porou matrix, allowing for the observation and order-of-magnitude analysis of the fluid dynamics associated with phase change heat transfer governed by double diffusion. Thermal and solutal gradients both affected the heat transfer and melting process. Thermal convection induced by thermal gradient promoted the melting of the upper section, while salt perticipated enhanced the melting of the lower section. The following conclusions were obtained from the study:

(1) During melting in the hypoeutectic side, a protruding melting region at the lower part of the melting interface, attributed to concentration stratification and the concentration-induced reduction in melting point. When melting in the hypereutectic side, the entire melting interface is downward curved due to the fact that salt stratification affects the flow pattern to a greater extent than thermal convection.

(2) The melting process is discernibly delineated into three distinct stages, with soil matrix influencing flow patterns, thereby impacting heat transfer and melting rates. By employing dimensionless analysis, a correlation expression for the Nusselt (Nu) number has been derived. The correlation equation reveals that the melting process is predominantly influenced by several key parameters, including the buoyancy ratio, Darcy number, sample height, thermal diffusivity, and Rayleigh number. These factors collectively govern the convective heat transfer characteristics during the melting process and provide valuable insights into the underlying physics of the system.

(3) Through order of magnitude analysis, in the first stage of melting, heat conduction is dominant and the $Nu$ number is mainly affected by the temperature difference, the thermal diffusivity of the solution and the height of the square cavity. In the second stage, $Nu$ number are positively related to $Ra_m$, incorprating the influence of permeability, buoyancy ratio and heating power. Larger $Ra_m$ indicates a higher intensity of the thermal natural convection. Also, the relationship between the $Nu$ number and SteFo becomes complex. In the thrid stage, the Nu number is independent of SteFo and depends only on the permeability of the porous medium, the density of heat flow and the buoyancy ratio.

Based on the expressions obtained from the order-of-magnitude analysis, it is possible to estimate the time scale required to melt a given amount of salty ice for different input powers and using different heights of heater, as well as for different thermal conductivities, permeabilities, and salinities, which may provide insights for the related applications.